\documentclass[12pt,preprint]{aastex}

\def\kms{km s$^{-1}$}

\def\mic{{$\mu$m}}

\def\h2o{H$_2$O}

\def\aple{$\mathrel{\hbox{\rlap{\hbox{\lower4pt\hbox{$\sim$}}}\hbox{$<$}}}$}

\def\apge{$\mathrel{\hbox{\rlap{\hbox{\lower4pt\hbox{$\sim$}}}\hbox{$>$}}}$}

\begin{document}

\title{Accretion Signatures from Massive Young Stellar
Objects\footnote{Based on observations obtained at the Gemini
Observatory, which is operated by the Association of Universities for
Research in Astronomy, Inc., under a cooperative agreement with the
NSF on behalf of the Gemini partnership: the National Science
Foundation (United States), the Particle Physics and Astronomy
Research Council (United Kingdom), the National Research Council
(Canada), CONICYT (Chile), the Australian Research Council
(Australia), CNPq (Brazil) and CONICET (Argentina)}}

\author{R. D. Blum}
\affil{Cerro Tololo Interamerican
Observatory, Casilla 603, La Serena, Chile\\ rblum@ctio.noao.edu}

\author{C. L. Barbosa \& A. Damineli}

\affil{Instituto de Astronomia, Geof\'{\i}sica e Ci\^encias
Atmosf\'ericas - USP, Rua do Mat\~ao, 1226 Cid. Universit\'aria S\~ao
Paulo-SP CEP 05508-900, S\~ao Paulo, Brasil \\cassio@astro.iag.usp.br,
damineli@astro.iag.usp.br}

\author{P. S. Conti} \affil{JILA, University of Colorado\\Campus Box
440, Boulder, CO, 80309\\pconti@jila.colorado.edu}

\author{S. Ridgway} \affil{National Optical Astronomy Observatory,
950 N. Cherry Street, Tucson, AZ, 85719\\sridgway@noao.edu}

\begin{abstract}

High resolution ($\lambda/\Delta\lambda =$ 50,000) $K-$band spectra of
massive, embedded, young stellar objects are presented. The present
sample consists of four massive young stars located in nascent
clusters powering Galactic giant H~II regions. Emission in the 2.3
\mic \ 2--0 vibrational--rotational bandhead of CO is observed. A
range of velocity broadened profiles seen in three of the objects is
consistent with the emission arising from a circumstellar disk seen at
various inclination angles. Br$\gamma$ spectra of the same spectral
and spatial resolution are also presented which support an accretion
disk or torus model for massive stars. In the fourth object, Br
emission suggesting a rotating torus is observed, but the CO profile
is narrow, indicating that there may be different CO emission
mechanisms in massive stars and this is consistent with earlier
observations of the BN object and MWC~349. To--date, only young
massive stars of late O or early B types have been identified with
clear accretion disk signatures in such embedded clusters. Often such
stars are found in the presence of other more massive stars which are
revealed by their photospheric spectra but which exhibit no disk
signatures.  This suggests the timescale for dissipating their disks
is much faster than the less massive OB stars or that the most massive
stars do not form with accretion disks.

\end{abstract}


\section{INTRODUCTION}

How do massive stars form? Simple expectations based on the observed
masses and life times of massive stars dictate that the timescale for
formation must be much shorter than for low mass stars and that the
average mass accretion rates must be high ($\sim$ 10$^{-4}$ M$_\odot$
yr$^{-1}$). Detailed calculations suggest the formation timescale is
approximately 10$^5$ yr \citep{mt03}. In contrast to low mass stars
which are more numerous and whose formation proceeds by distinct and
identifiable phases of longer timescales, high mass stars are rare,
their formation rapid and obscured by an overlying envelope. It is
also unavoidable that a massive core will form and begin evolving
(converting H to He) while the larger protostar is still accreting
material. This makes for theoretical complexity as well as
observational difficulty.

Despite these difficulties, progress is being made on both theoretical
and observational fronts; see the recent reviews by \cite{gl99},
\citet{kcchw00}, and \citet{ec02}. A number of groups have developed
accretion models for massive stars: \citet{bm01}, \citet{ys02}, and
\citet{mt03}. Observational evidence for accretion disks has been
accumulating through mm and molecular line measurements for a few
highly embedded objects \citep{gl99,sk99}. While massive protostars,
these objects are not of the most massive type but tend to be mid to
early B stars. Methanol masers have been used to trace sites of
massive star formation, and have been proposed to trace circumstellar
accretion disks \citep{norris98,phillips98} though there are alternate
explanations \citep{walsh98}. A recent survey of these objects for
molecular outflows (H$_2$ emission), which are expected to be
associated with the putative disks, also argues against the disk
origin for methanol emission \citep{deb03}.

Another possible formation mechanism for massive stars is through
mergers of lower mass protostars in a dense cluster
environment. Recent simulations have been successful in producing
massive stars through this mechanism \citep{bbz98,sph00,bb02}. The
details of the \citet{bb02} simulation show that the stellar cluster
density evolves rapidly at the end of the formation process reaching
densities suitable for mergers. This suggests it will be difficult to
catch a massive star ``in the act'' and so direct observational
confirmation of merger formation of the most massive stars may be very
difficult. On the other hand, lack of detection of accretion
signatures for the most massive stars can be used to infer the mass at
which mergers would become the dominant channel of formation. This
would also require substantial observational effort.

Obviously, investigations of the earliest phases of massive star
formation have centered around longer wavelengths which penetrate the
large column depth of material surrounding massive protostars. But
current angular resolutions and sensitivities achievable restrict
study to the relatively few nearby sources (the ALMA telescope will
help to rectify this situation!). In the present work we take
advantage of the fact that newly formed massive stars, or massive
young stellar objects (MYSOs), while at the tail end of the star birth
process, can be used to search for evidence of remnant accretion
disks. Our ongoing survey of Galactic giant H~II (GHII) regions in the
near infrared \citep{bdc99,bcd00,bdc01,cb01,fig02,fig04} combined with
the work of \citet{hhc97} has led to the discovery of a sample of
MYSOs with potential disk signatures which we describe below. These
candidates have been observed at high angular and spectral resolution
in a variety of $K-$ band emission lines. In this paper, we discuss
the observations of Br$\gamma$ and (in more detail) CO 2.3 \mic \
emission. The latter can be particularly useful in studying the
circumstellar geometry of MYSOs as has been shown to be the case for
low mass YSOs \citep{carr89,carr93,chan93,chan95,naj96}.

\section{OBSERVATIONS AND DATA REDUCTION}

\subsection{Sample Selection}

Our sample of stars consists of $K-$band bright and $H-K$ red objects
found in our imaging survey of GHII regions (Table~1) and the study of
M17 by \citet{hhc97}. These objects are often the brightest near
infrared sources in the associated clusters and have colors which
indicate strong excess emission. None of these stars have $K-$band
photospheric features present in low resolution spectra, but all are
probably late O or early B type stars based on other (primarily
luminosity) arguments; see the references listed in Table~1. The
complete sample of such massive stars still enshrouded by their birth
material and for which at least low resolution spectra are available
is currently about 10 (our survey is still on--going). In this paper,
we present high spectroscopic resolution data for four of the stars
which show CO 2.3 \mic \ 2--0 vibrational rotational bandhead emission
in their low resolution spectra.

\subsection{Gemini/Phoenix Observations}

The target spectra were obtained with the NOAO Phoenix spectrometer on
the Gemini South 8m telescope in 2002 and 2003 (see Table~1). Phoenix
is described in detail by \citet{hinkle03}. In all cases the data were
obtained with the four pixel (0.32$''$) wide slit giving a resolution
of $\lambda/\Delta\lambda \approx$ 50,000. The resolution element is
over--sampled in this mode giving approximately five pixels per
element. The Phoenix pixel scale on Gemini South is approximately
0.09$''$ pix$^{-1}$ resulting in a slit length of about 15$''$.

Spectra were obtained in both photometric and non--photometric
conditions and for both Br$\gamma$ (2.1661 \mic, all wavelengths in
this paper are quoted in vacuum) and the first overtone CO 2--0
bandhead (2.2935 \mic). For one star, NGC3576 \#48, two grating
settings near the CO line were obtained. All other stars have data for
one grating setting only. The seeing (at $K$) ranged from
approximately 0.4$''$ to 1.3$''$. Typically it was close to 0.5$''$,
with the exception of the Br$\gamma$ observations for NGC3576 \#48
(1$''$ -- 1.3$''$). A similar spectrum of NGC3576 \#48 at the position
of Br$\alpha$ was also obtained on the same night.

The two--dimensional long slit images were reduced with the
IRAF\footnote{ IRAF is distributed by the National Optical Astronomy
Observatories, which are operated by the Association of Universities
for Research in Astronomy, Inc., under cooperative agreement with the
National Science Foundation.} data reduction package. Each image was
divided by a flat field image obtained with the Gemini GCAL unit and
then subtracted by a sky image. The Br data had independent
sky frames observed off the targets since the emission often filled
the entire slit. The CO sky frames were obtained by nodding the target
along the slit by about half a slit length (i.e. the target was always
on the slit) since the CO emission was spatially unresolved and
centered on the point source. The Br images were obtained with
N--S and/or E--W slit orientation on the sky, while the CO data always
had a E--W orientation.

After extracting one--dimensional spectra (with $\pm$ 2 to 5 pixels
along the slit $\sim$ 0.2$''$ to 0.5$''$), the spectra were divided by
similar spectra obtained on atmospheric standard stars. For the Br
spectra, intrinsic absorption in the standards was removed by fitting
the absorption features dividing the fit into the original
spectra. The final ratioed spectra were wavelength calibrated using
telluric absorption lines measured off the electronic version of the
NOAO Arcturus atlas \citep{hinkle95}. A linear fit to the telluric
line positions gives a typical uncertainty of about $\pm$ 0.5 to 1.0
pixel (i.e. 1/10th to 1/5th of a resolution element), or $\pm$ 0.7 to
1.3 \kms.

\section{RESULTS}

\subsection{CO Emission}

\begin{figure}
\plotone{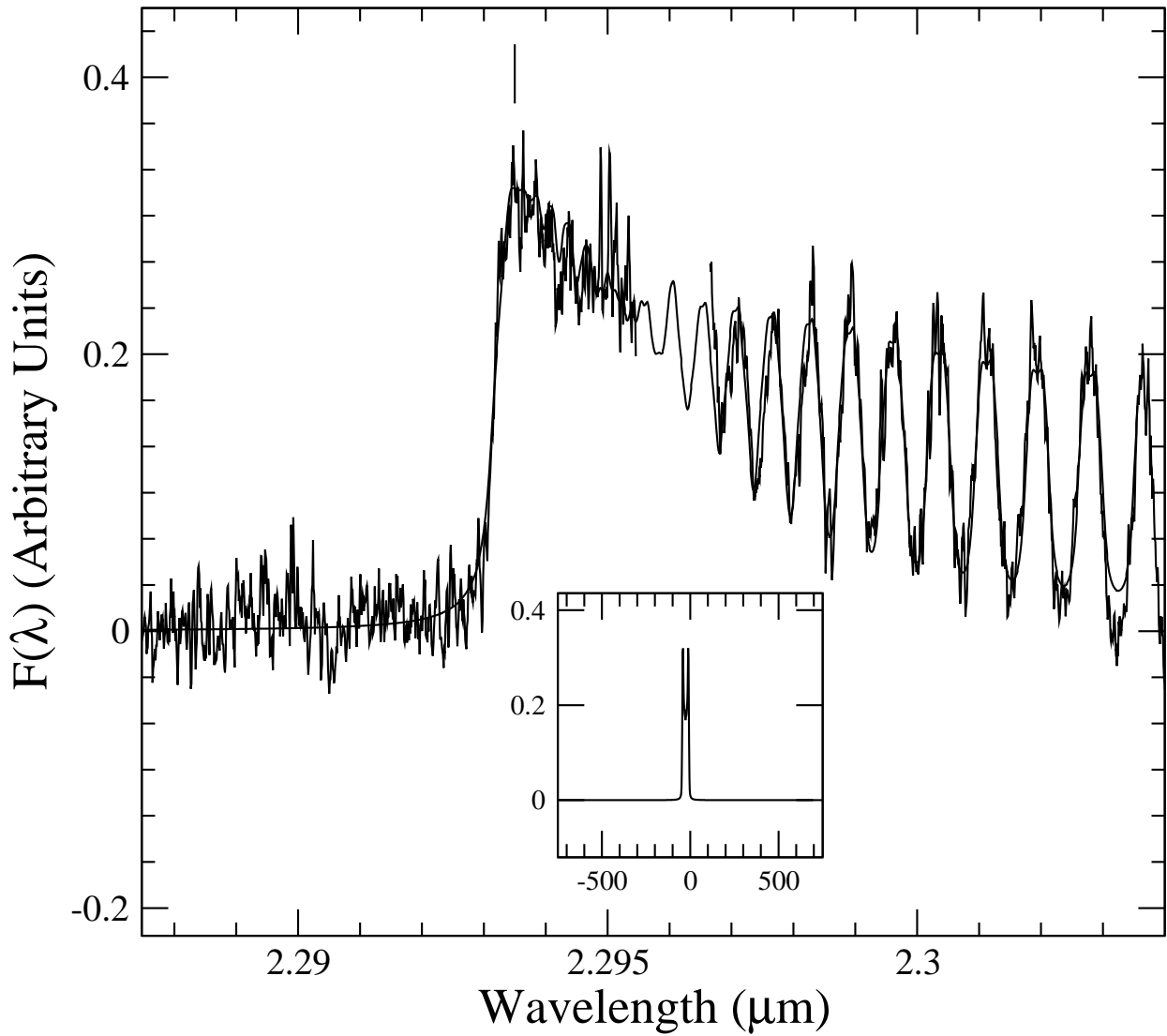} \figcaption{The CO 2-0 first overtone
rotational--vibrational bandhead for source \#48 in NGC3576. The
smooth curve is a model for the emission profile arising from a
Keplarian disk (see text and Table~2). The spectrum is a sum of two
grating settings. The region near 2.296 \mic \ is the interface
between the two, and the array has a number of bad pixel rows on one
end of the detector making the spectrum here unreliable, and it has been 
masked out in this figure. The short
vertical line marks the vacuum rest wavelength of the bandhead (2.2935
\mic), and the inset shows the emission--line profile for a single 
line in the bandhead ($v =$ 2--0, $J=$51--50). The $x$ axis of the inset
is in \kms \ and is one Phoenix grating setting in extent.\label{co48}}
\end{figure}

\begin{figure}
\plotone{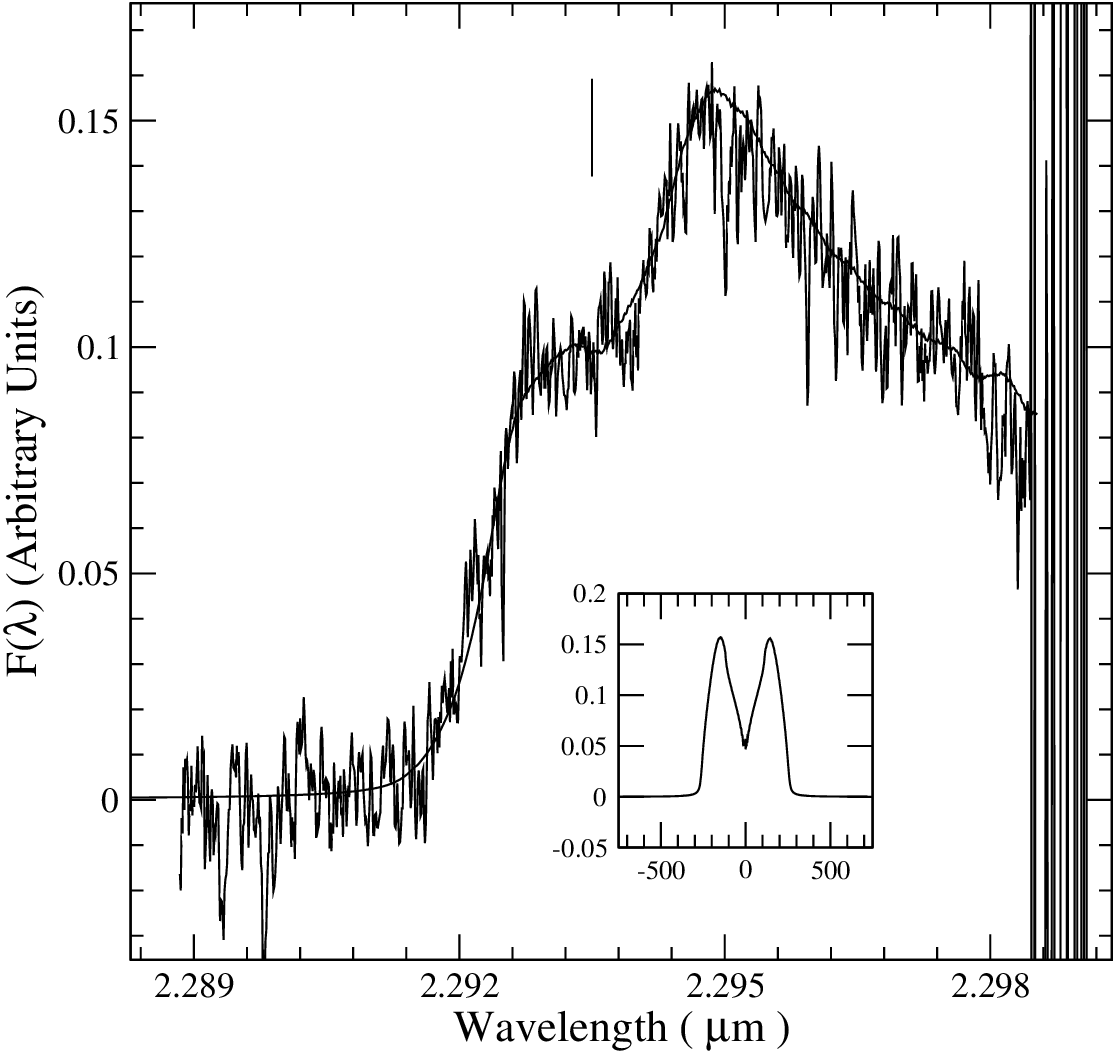} \figcaption{Same as Figure~\ref{co48} but for
source 268 in M17. \label{co268}}
\end{figure}

\begin{figure}
\plotone{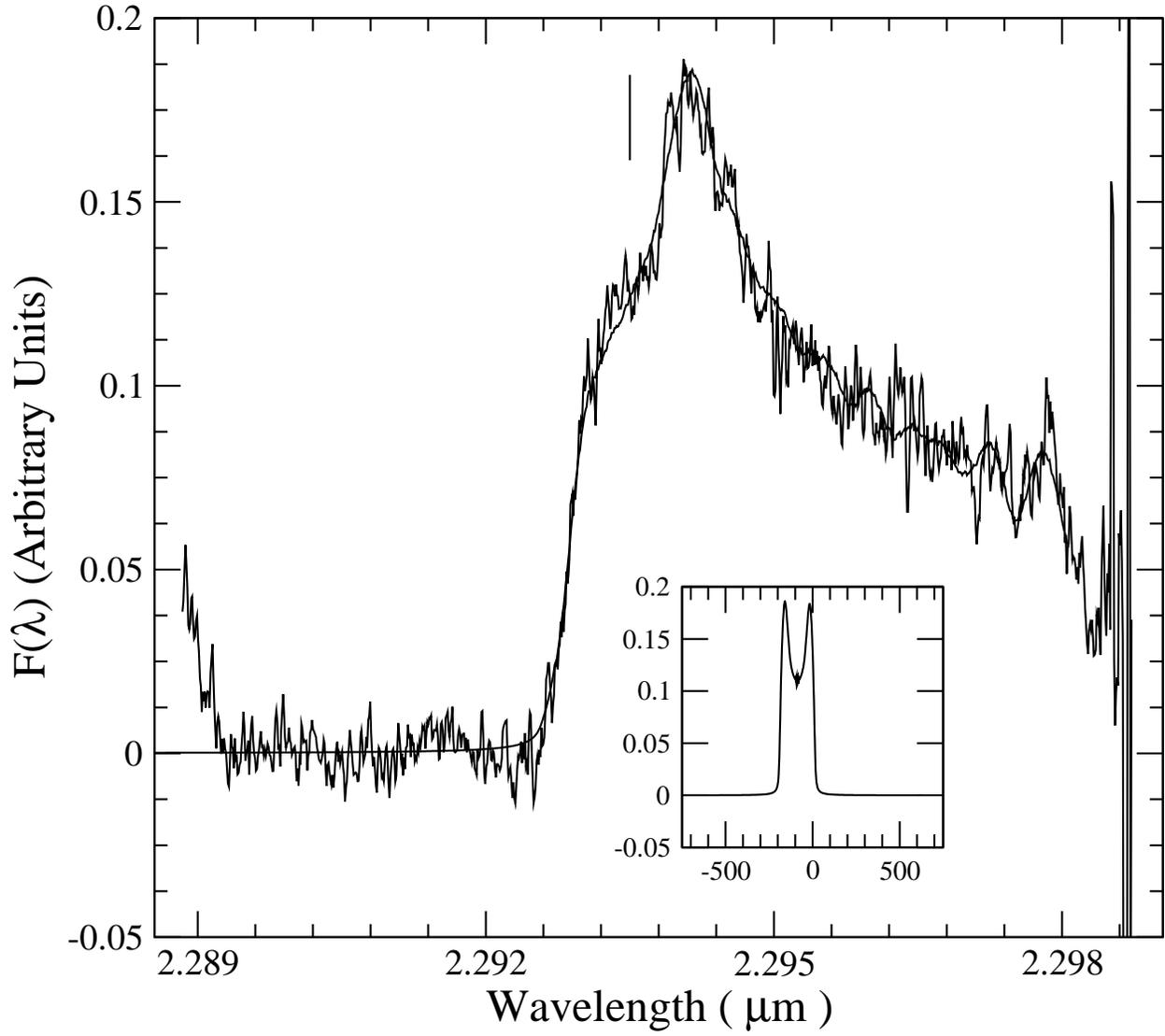} \figcaption{Same as Figure~\ref{co48} but for
source 275 in M17. \label{co275}}
\end{figure}

\begin{figure}
\plotone{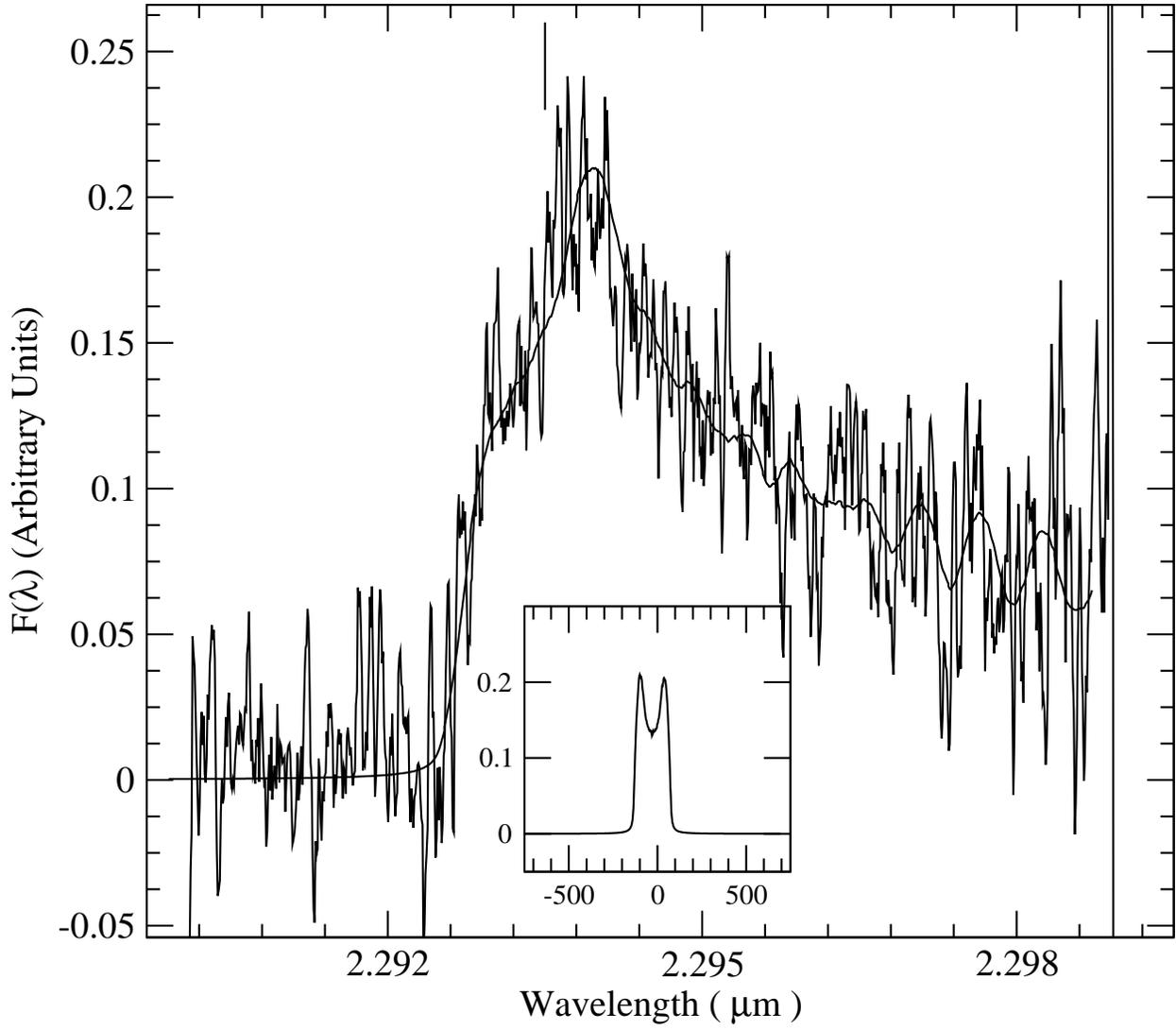} \figcaption{Same as Figure~\ref{co48} but for
source \#4 in G333.1--0.4. \label{co4}}
\end{figure}

The individual spectra are shown in Figures~\ref{co48}, \ref{co268},
\ref{co275}, and \ref{co4}. The spectra have been binned by a factor
of three to improve the signal to noise and the continuum has been
subtracted. These spectra show a range of profiles from narrow
emission (NGC3576 \#48, Figure~\ref{co48}) to very broad emission
(M17--268, Figure~\ref{co268}).

The shape of the CO emission profile in these spectra is the same as
that expected for a rotating Keplarian disk. In particular,
Figure~\ref{co268} shows a profile for M17--268 which is a good match
to the profile for the lower mass stars analyzed by \citet{carr93},
\citet{chan95}, and \citet{naj96}. As discussed by \citet{carr93}, the
pronounced blue--shifted shoulder and redshifted emission peak are
well fit by a rotating disk.

Following the procedure outlined by \citet{kraus00}, we fit the
emission profiles of each of the four massive stars in our
sample. These models use power--laws in the disk temperature and
surface density distributions with optical depth and flux in the CO
transitions calculated in each cell of a radial plus azimuthal
coordinate grid. The emission from each cell is added to the final
spectrum according to its rotational velocity and inclination. Within
each cell, an intrinsic line profile is assumed. Following
\citet{naj96}, we have adopted a Lorentzian profile (which allows for
lower column densities; see Najita et al. 1996), though Gaussian
profiles of larger width can also fit the data. In the range of
wavelengths covered by the present Phoenix spectra, the Hydrogen Pfund
lines are expected to be weak \citep[see][]{kraus00}.

We adopt a temperature of 5000 K (above which CO should be
dissociated) at the inner radius of the CO emission zone. The column
of material emitting is assumed to be geometrically thin with uniform
temperature and density in the direction normal to the disk so that
the optical depth through any cell in the disk is just the absorption
coefficient for a single CO molecule times the total surface density
(cm$^{-2}$). We do not fit the continuum, only the line
emission. Similar models which consider a two temperature disk
(i.e. the structure normal to the disk plane has two components) with
an emission component overlying a continuum component have been
constructed by \citet{naj96}. For the purpose of the exploring the
kinematic signatures of the disks, a single emission component is
sufficient.

The observed profiles were fit by systematically varying the free
parameters. To obtain a rough fit, the parameters were varied and the
results compared by eye to the observed spectra.  The free parameters
are the temperature distribution exponent, $p$, the density
distribution exponent, $q$, the total column density at the inner
radius for emission, $N_{\rm CO}$, vsini at the inner radius, the
systemic velocity of the system, the width of the intrinsic line
profile, and the ratio of outer to inner radius, $\beta$. In practice,
the fits are not very sensitive to the value of $p$ and it was fixed
for all models to 0.5.  Good fits were obtained in all cases with the
adopted profile width of 1 \kms \ (see above).  Once the initial
parameters were obtained by eye, we improved the fits using a simple
grid search method, minimizing Chi square as described in
\citet{bev92}.  For the Chi square minimization, the values of $q$,
$\beta$, vsini, and $N_{\rm CO}$ were varied in turn and the minimum
Chi square found for each independently. The procedure was iterated
one time with the reduced Chi square changing by less than 1$\%$ in
all cases. The reduced Chi square values ranged from 0.75 to 1.4.  The
parameters are summarized in Table~2, and the best fits obtained by a
systematic variation of the free parameters are shown as the smooth
black lines in each of Figures~~\ref{co48}, \ref{co268}, \ref{co275},
and \ref{co4}.  The formal one sigma errors on the fitted parameters
are quite small, and the models clearly fit the data well as seen in
Figures~~\ref{co48}, \ref{co268}, \ref{co275}, and \ref{co4}. However,
a more detailed treatment including more CO transitions would be
required to definitively explore the physical state of these
disks. The goal of the present simplified treatment is to show that
plausible kinematic models of rotating disks fit the CO profiles well.

The systemic velocity was determined by cross--correlation of the model 
and observed profiles using the IRAF "rv" package. The cross--correlation peak 
center region was fit in each case with a Gaussian, and the formal error in 
its center is less than one pixel in all cases. We have added to this 
uncertainty (in quadrature) the typical 
uncertainty cited above (\S2) from the wavelength solution. 

For NGC3576 \#48, there is essentially an additional parameter which
is the relative flux level between the two grating positions. The
region where the two pieces are spliced together is also unreliable
due to bad pixel rows at the end of the detector. There appears to be
a slight over--subtraction of the continuum at the red end.

\subsection{Br$\gamma$ Emission}

\begin{figure}
\plotone{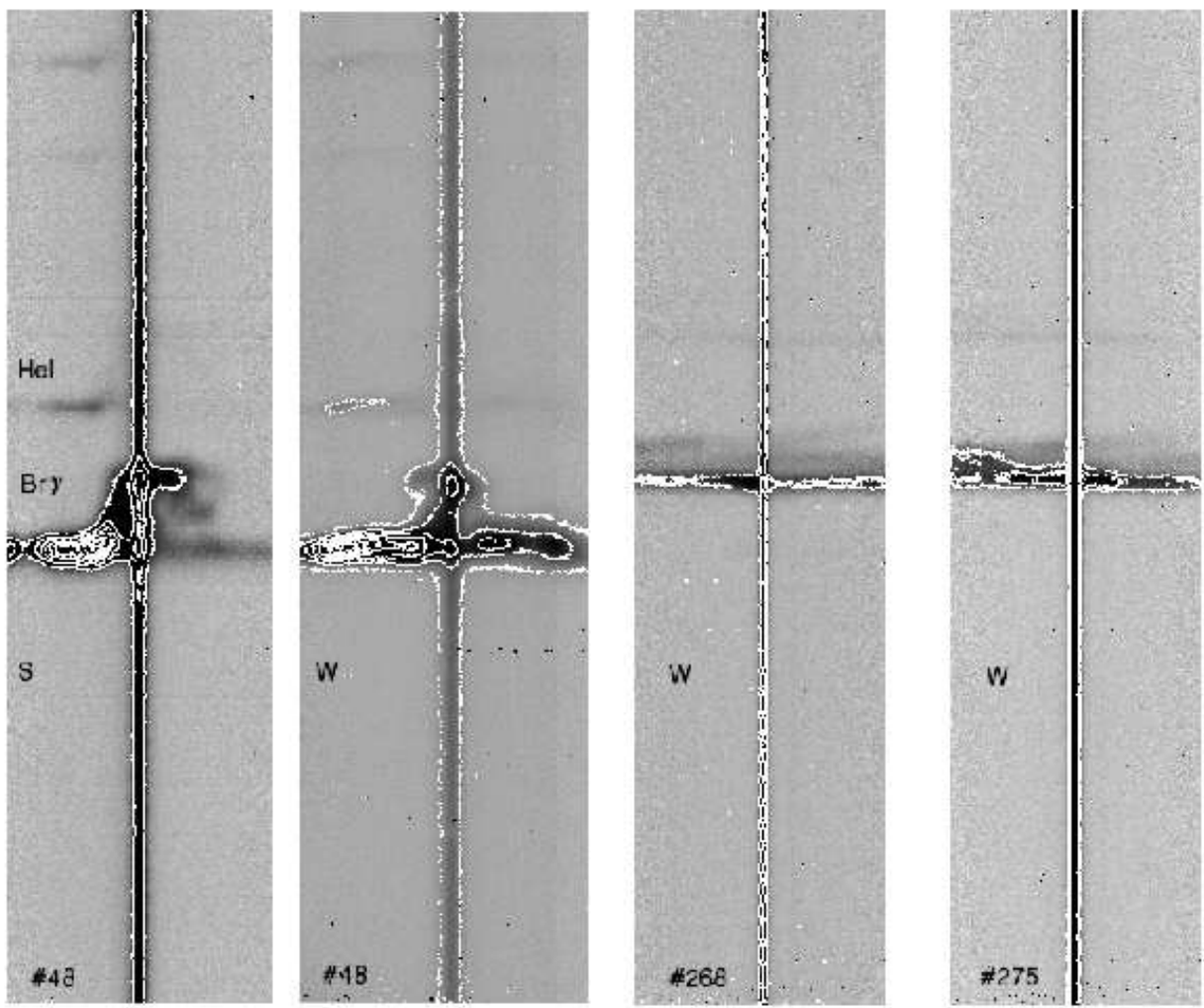} \figcaption{Two dimensional long slit images in the
Br$\gamma$ grating position for the NGC3576 \#48, and the M17 sources:
{\it far left}, N--S slit for NGC3576 \#48, {\it middle left}, E--W
slit for NGC3576 \#48, {\it middle right}, M17--268, {\it far right},
M17--275. West and South are to the {\it left} in these images. 
The spatial dimension is along the horizontal axis in each image
and covers approximately 20$''$--24$''$ along the slit. The images
are combinations of the individual dither positions. 
The vertical axis is the dispersion (or spectral) dimension,
and its velocity extent is approximately 1500 \kms \ for each image. 
He~I 7--4 emission is present in each
case to the blue of Br$\gamma$ (both lines marked in the \#48 N-S
image). The HeI emission has a compact component (i.e. superimposed on the
continuum source) just as does the
Br$\gamma$ emission; see Figure~\ref{brgs48} and text.
The contours (derived from the same images) 
are overlaid on the gray scale to give an indication of the
relative strength of the emission along the slit and compared to the continuum 
source.\label{2d}}
\end{figure}

Unlike CO emission which is spatially unresolved, the Br$\gamma$
emission is typically extended along the slit in our 2--D images. The
ionization can be due to the central source and/or other sources.  In
Figure~\ref{2d}, we display the long slit images for NGC3576 \#48,
M17--268, and M17--275. The spatial dimension runs along the
horizontal axis, and the spectral or dispersion dimension runs among
the vertical axis.  These images are rectified (shifted so the
spectrum runs vertically in the image), spatially shifted, and
combined images which cover up to 24$''$ along the slit (a single
exposure corresponds to a 15$''$ slit). Contours derived from the same
images are overlaid on the gray scale to indicate the relative
strength of emission along the slit and compared to the continuum
source. The gray scale itself has been stretched to show faint detail
in the emission morphology.

The emission away from the central source of NGC3576 \#48 is intense,
peaking $\sim$ 4$''$ (0.05 pc, for $d$ $=$ 2800 pc, de Pree,
Nysewander, \& Goss 1999) south of the continuum source in the N--S
image (see Figure~\ref{2d}). There is a somewhat lower ``plateau'' of
emission out to 6.5$''$ (0.09 pc) south which then decreases sharply
to a relatively low value at 9.4$''$ (0.13 pc) south. Beyond this
point, the emission may increase again. To the north, the emission is
much less intense, only a maximum of 6$\%$ of the peak; again, a clear
cuttoff is apparent at about 8$''$ (0.11 pc) north. An E--W slit
spectrum was also taken. In the E--W case, the emission is much
stronger on the west side of the continuum source, peaking at 9$''$
(0.12 pc) west (see Figure~\ref{2d}). The emission centered on the
source in both slit orientations shows a double peak and is also
strong, about 34$\%$ and 43$\%$ of the peak in the E--W and N--S
orientations, respectively. The two velocity extremes have a
peak--to--peak difference of about 85 \kms $\pm$ 1.5 \kms in each
case, and the mean velocity of this compact or ``torus'' component
appears blue--shifted from the bright off source emission to the south
and west. The compact emission is interpreted as a rotating torus
below.

\begin{figure}
\plotone{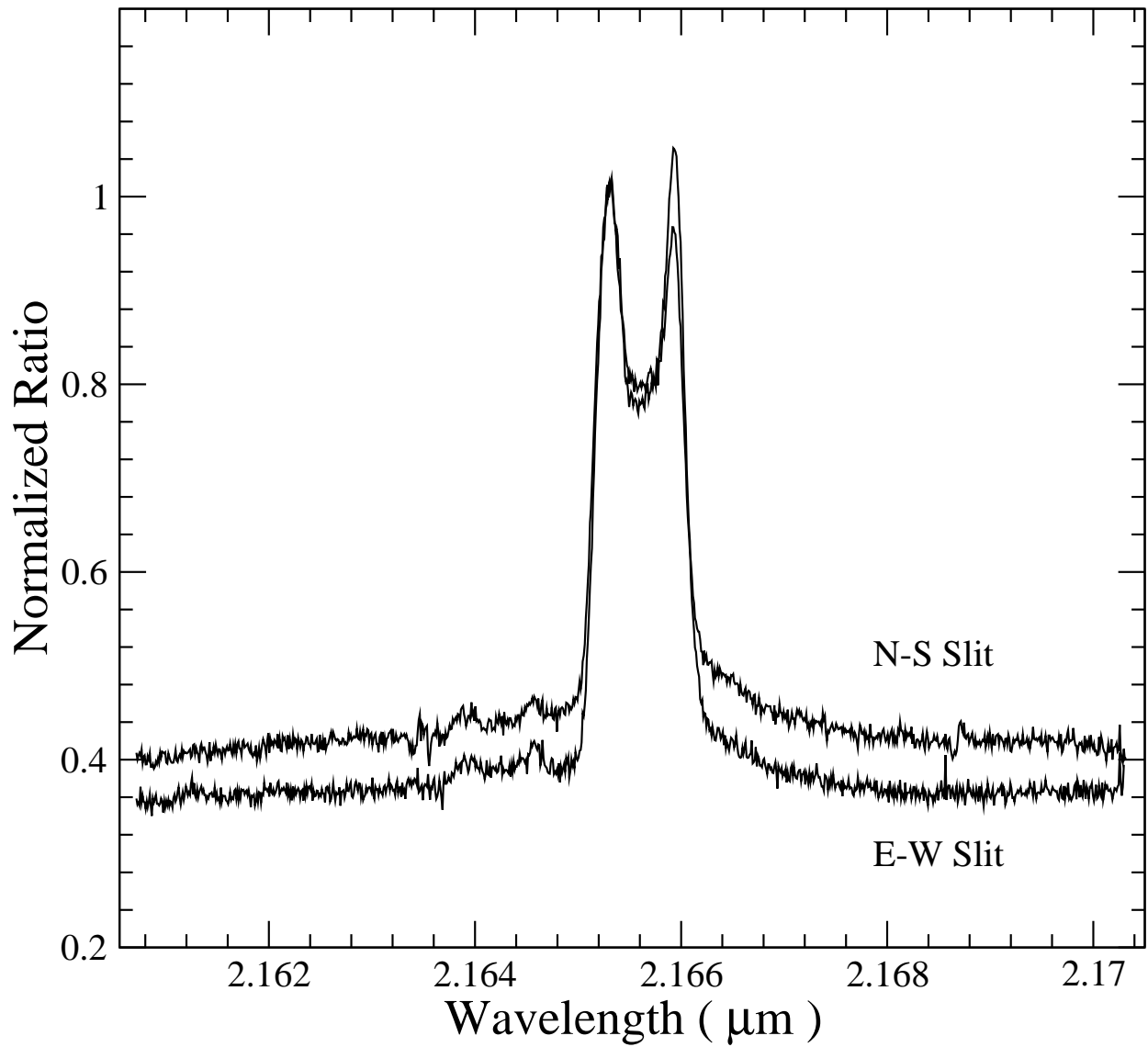} \figcaption{Br$\gamma$ emission for source \#48
in NGC3576. The two curves represent the emission seen through the
Phoenix 0.32$''$ slit and summed over pixels corresponding to 0.9$''$
along the slit spatial dimension. No nebular background is subtracted from
these spectra because it is strong and variable away from the
source. The two spectra were obtained with different slit
orientations. The lower spectrum had an E--W slit while the upper had
N--S.\label{brgs48}}
\end{figure}

Br$\gamma$ spectra for NGC3576 \#48, M17--268, and M17--275 are shown
in Figures~\ref{brgs48}, \ref{brg268}, and \ref{brg275}, respectively
(we were unable to obtain a spectrum of G333.1--0.4 \#4). These
spectra were extracted from the original individual frames, not the
2--D images shown in Figure~\ref{2d}. Each spectrum is an extraction
over the central $\sim$ 0.3 to 1$''$ by 0.32$''$ slit width. The
emission in the NGC3576 \#48 slits is highly variable, and this
precluded using background apertures to isolate the emission close to
the stellar source. However, the emission is strong at the position of
the source and should therefore be representative of the circumstellar
component. This is supported by the similarity of the spectrum
extracted from the N--S and E--W slits. Because the emission in the
M17 sources is more uniform across the sources, background apertures
could be used.  Figure~\ref{brg268} and \ref{brg275} show the total
emission seen through the Phoenix slit and the spatially compact
component obtained by subtracting the uniform component using 0.9$''$
wide apertures parallel to the continuum source and located at $\pm$
1.3$''$ away. In each case, a relatively strong but narrow emission
component is removed leaving a much broader component which arises
from a region close to the central star.

Table~3 summarizes the Br$\gamma$ velocities for the central compact
emission component(s). For each source, there is a mean Br$\gamma$
velocity, fullwidth, and peak--to--peak velocity difference (M17--268
and NGC3576 \#48). These parameters were determined by least squares
fitting to Lorentzian or Gaussian profiles. The formal uncertainties
are less than 1 pixel. In practice, the setting of the continuum
dominates the uncertainty in the fit. For a range of continuum
choices, the fits suggest centroids which vary by approximately $\pm$
two pixels. The typical residual in the position of lines used in the
wavelength solution (one pixel) is added in quadrature to this
value. The FWHM are similarly affected by the continuum choice. The
peak--to--peak separations for NGC3576\#48 and M17--268 were
determined by calculating the centroids of the individual peaks. The
formal uncertainty is taken conservatively as the typical residual in
the wavelength solution (one pixel).

He~I 7--4 emission is present in each case to the blue of Br$\gamma$
(about 1/4 of the way from the top of each image). The rest wavelength
of the emission is 2.1647 \mic \ \citep{fig97} and its presence
clearly establishes the ``hot'' nature of the central sources.  The
He~I peak--to--peak velocity is 93 $\pm$ 1.5 \kms \ compared to 85
$\pm$ 1.5 \kms \ for Br $\gamma$ in NGC3576 \#48 (the mean velocities
are equal within the one sigma uncertainties) which suggests the He~I
emission arises from a region slightly closer to the central source in
models where the emission arises due to rotation (see below).

\section{DISCUSSION}

\subsection{The Origin of the CO Emission in MYSOs}

\citet{sco83} presented high resolution spectra of the
Becklin--Neugebauer (BN) object which show many similarities to the
spectra presented here. In particular, BN exhibits strong Brackett
series emission and CO 2.3 \mic \ emission. BN is also thought to be a
heavily shrouded early B--type star. In their analysis of BN,
\citet{sco83} preferred an interpretation for the observed narrow CO
emission which was not produced in a rotating circumstellar
disk. Except for the case where BN was viewed nearly face on, the
small required area, and hence radius, of the bandhead emission region
requires a high rotational velocity which is not
observed. \citet{sco83} proposed a shock model for the origin of the
CO emission instead, placing this physically narrow region much
farther from the star.

\citet{carr93} and \citet{chan95} modeled the CO emission from other
young stars, including several with masses up to $\sim$ 25 M$_\odot$,
and found excellent fits for a disk origin. \citet{chan95} also
compared their data to wind models, but found that the disk models
generally fit better. The CO profiles in our spectra of M17--268,
M17--275, and G333.1--0.4 \#4 are very well fit by emission arising in
a Keplarian rotating disk with vsini between about 100 \kms \ and 260
\kms. For the estimated masses given in Table~2, this requires the CO
emission to arise at radii less than 1 AU from the central star in all
cases. Each has a profile which exhibits the characteristic shape of
CO emission produced by a Keplarian disk \citep{carr93,chan95},
including most notably, the blue shoulder and redshifted emission
peak.  The case for massive star formation of late O and early B stars
by accretion seems quite solid. NGC3576\#48, on the other hand, shows
a very narrow profile which could be consistent with either a disk
seen nearly face--on or with some other geometry such as a wind or
inflow/outflow.

Is BN a special case? Besides our spectrum of NGC3576 \#48, a similar
spectrum of the 2.3 \mic \ first overtone emission is presented by
\citet{kraus00} for the MYSO MWC~349. The narrow emission is difficult
to fit with a disk model or a combination disk plus wind. This is
because \citet{kraus00} fix the inclination of the disk to be nearly
edge on, based on independent measurements of ionized hydrogen lines
and radio continuum \citep{rod94}. For such a geometry, the narrow CO
emission profile results in an emission region which is far from the
central source where the temperature and surface density are expected
to be much lower based on observations of the associated dust disk.

The observations presented here for NGC3576 \#48 also show narrow CO
emission (Figure~\ref{co48}). In this case, the observed profile could
be due to inclination effects. However, the Br$\gamma$ emission
depicted in Figures~\ref{2d} and \ref{brgs48} shows a double peaked
profile with a velocity separation between peaks of 85 \kms. This
profile could arise from an expanding shell or rotating torus. Similar
profiles occur in MWC~349 \citep{hs86} for a variety of lines
including Br$\gamma$. The Br$\gamma$ emission in MWC~349 has a
peak--to--peak velocity difference of about 50 \kms. In the case of
MWC~349, \citet{hs86} interpret the emission as arising from the
ionized face of a circumstellar disk. \citet{barb03} developed a
flared torus model for NGC3576 \#48 in order to explain the
differences in the observed spectral energy distributions of NGC3576
\#48 and several other sources in the same cluster.

In both NGC3576 \#48 and MWC~349, a natural expectation is that the CO
emission would be produced in the inner warm, neutral component of
such a disk or torus. \citet{kraus00} have suggested a model which
qualitatively reconciles this expectation for MWC~349 with the
observed narrow CO profile. In their model, the large scale
circumstellar geometry is taken as a bulging disk. The bulged disk
shadows much of the CO emitting region in the inner part of the disk
(for inclinations nearly edge--on) allowing only a view toward a small
sector which limits the range of observed velocities (see Figure~17 of
Kraus et al. 2000). Material from the ionized portion of the disk,
which is ablated off of it, fills the region above the disk and is
thus visible (consistent with the photo--ablation models of Hollenbach
et al. 1994). This material carries the kinematic signature of the
rotating disk it arises from \citep{hs86}.

While the model of \citet{kraus00} limits the range of velocities seen
in the CO bandhead emission, the mean velocity should be
unaffected. In the case of NGC3576 \#48, the mean CO velocity is
better matched to the bulk of the redshifted Br$\gamma$ emission, and
not that of the compact, double--peaked Br$\gamma$ emission presumably
arising in the torus (Table~2). The mean velocity of the CO and bulk
Br$\gamma$ emission are also in agreement with the radio recombination
line peak emission velocity for NGC3576 as given by \citet{depree99}.

\begin{figure}
\plotone{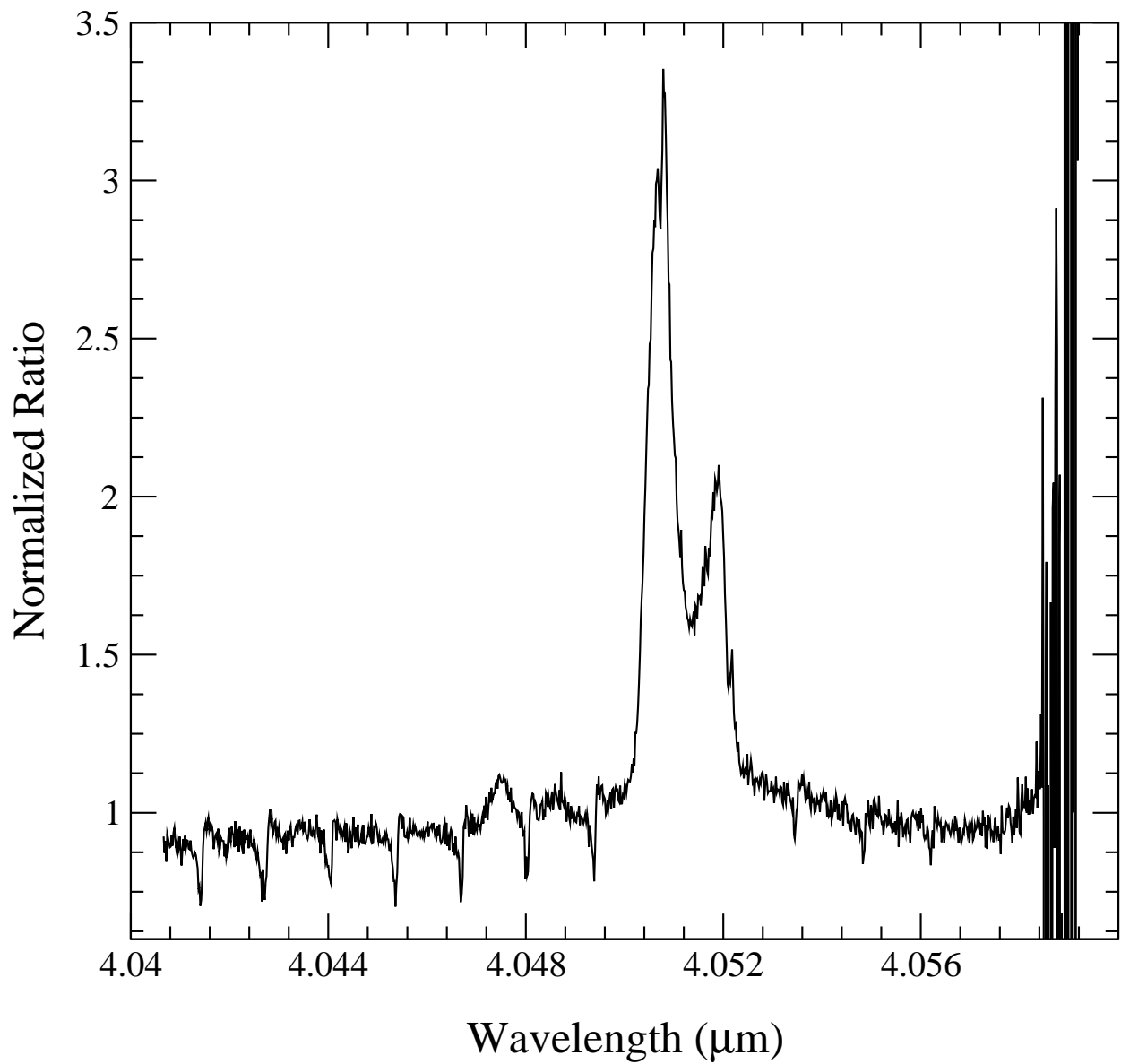} \figcaption{Same as Figure~\ref{brgs48} but for
Br$\alpha$ (4.052 \mic). No background has been subtracted. The slit
orientation is N--S. The blue shifted component is stronger than in
the Br$\gamma$ case because the longer wavelength emission is less
affected by the intervening dust; in this case, the emission arises
from the inner wall of the cavity surrounding NGC3576 \#48 (see
text). The blue peak, in particular is contaminated by the spatially
extended component of emission. The ``absorption'' features are due to
incomplete telluric correction. \label{balpha}}.
\end{figure}

\begin{figure}
\plotone{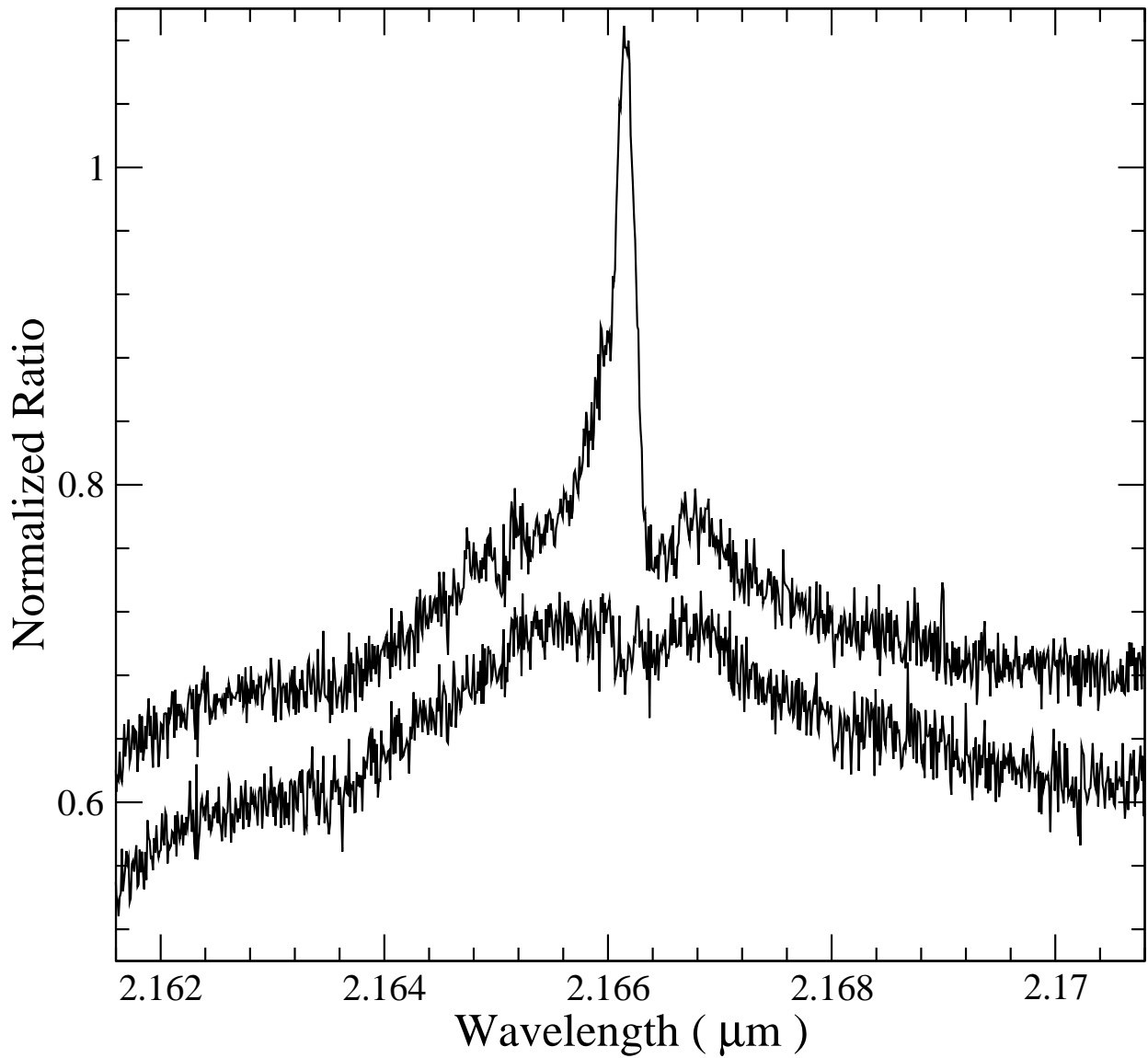}\figcaption{Br$\gamma$ emission for
M17--268. The upper curve represents the total emission seen through
the Phoenix 0.32$''$ slit and summed over pixels corresponding to
0.4$''$ along the slit spatial dimension. The lower curve is the same
except that the background has been subtracted from nearby ($\sim$
$\pm$ 1$''$) apertures. The two curves are nearly identical except for
the narrow blue emission peak and have been separated vertically for
clarity. The clean subtraction of the blue peak highlights the
uniformity of the background across the source. The background
subtracted spectrum has a broad, double--peaked profile consistent
with disk emission. The profile is approximately 450 \kms \ wide at
half the peak intensity. \label{brg268}}
\end{figure}

\begin{figure}
\plotone{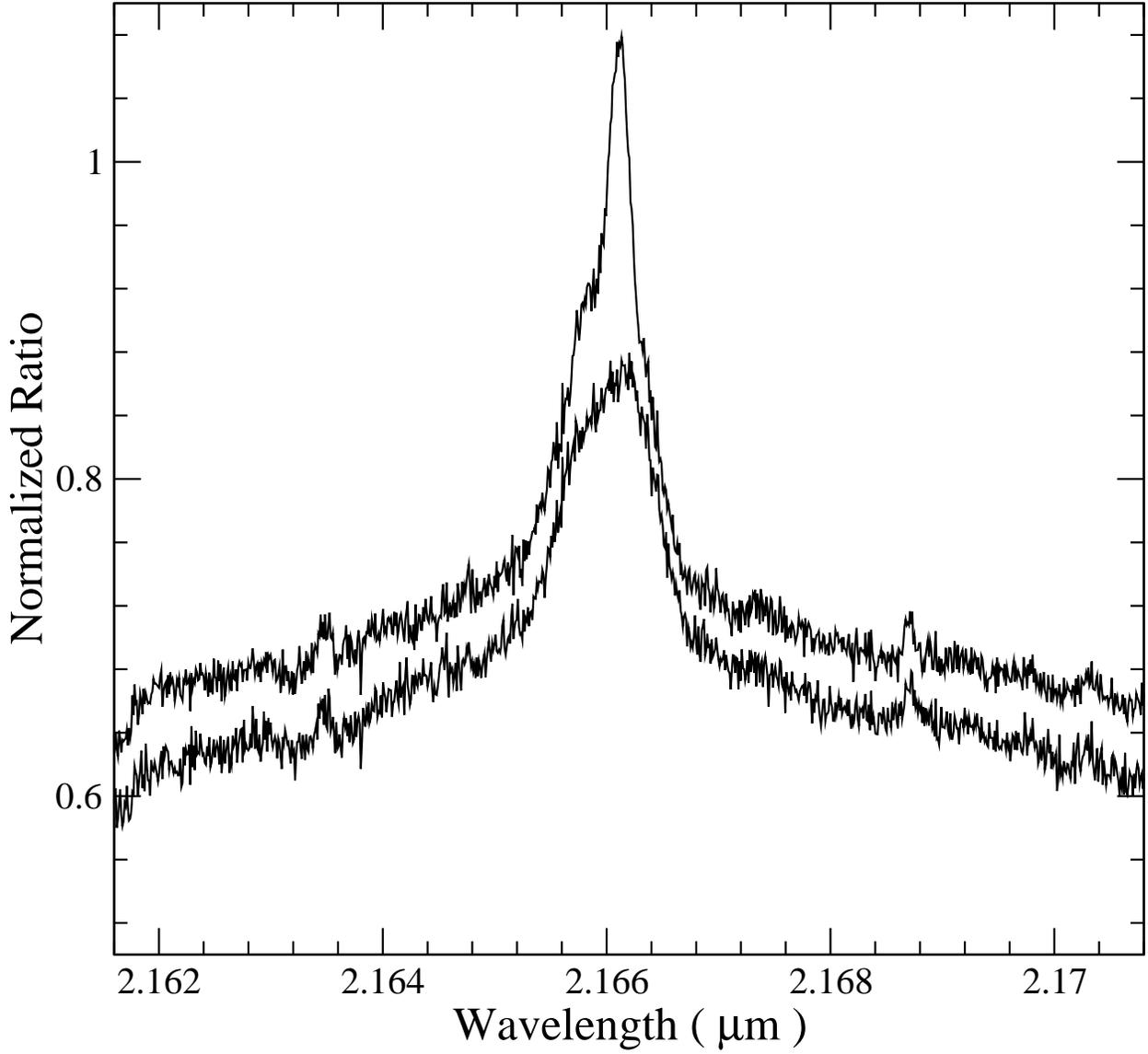} \figcaption{Same as Figure~\ref{brg268} but
for M17--275.\label{brg275}}
\end{figure}

With reference to the near infrared images of NGC3576 \#48 presented
by \citet{fig02}, the following picture emerges. The source is clearly
breaking out of its birth material. The images indicate that a ``cap''
of material running along the north side of the object provides a
column of obscuration in that direction along the line of sight. The
bulk of the cloud is to the south, but the immediate vicinity of
NGC3576 \#48 is cleared away in this direction providing a view inside
the cavity and beyond \#48; the cavity itself appears to be expanding
and breaking up due to the action of \#48 and other sources in the
cluster. \citet{barb03} suggest the main part of the cluster is
located to the SW of \#48 and obscured from our line of sight at these
wavelengths. This geometry is consistent with the velocities and
spatial distribution of the Br$\gamma$ emission seen in
Figure~\ref{2d}. The emission to the south ($far left$ panel) in
Figure~\ref{2d} is much stronger than to the north. The ``cap''
mentioned above shields the back side of the cavity which is evidently
being pushed away along our line of sight. The emission is not uniform
in the E--W direction, but much more emission is seen on both east and
west sides than in the north--south case. The blue shifted emission to
the north must originate on the inner wall of the ``cap.'' This part
of the cavity wall is being pushed toward us along the line of
sight. Further support for this geometry is provided by a Br$\alpha$
spectrum shown in Figure~\ref{balpha}. The blue shifted component of
the emission is much stronger relative to the case for Br$\gamma$. The
longer wavelength emission penetrates the ``cap'' more than the
shorter wavelength does. We interpret the compact emission in \#48 as
arising in a rotating torus, but some type of shell geometry is
possible as well. Indeed there is a weak continuous bridge of emission
connecting the torus to the bulk of the extended emission at a point a
few arcseconds to the south of the continuum source. The outer extent
of this shell feature can be seen in Figure~\ref{2d} as the lowest
contour plotted. There is a corresponding ``hole'' or lack of emission
which is not evident due to the stretch of the image.

The nebular and torus components are not as well separated in the
Br$\alpha$ case. Since the CO mean velocity is clearly redshifted to
roughly the same velocity as the bulk of the Br$\gamma$ emission and
not the mean velocity of the compact component, it appears the CO
emission is not arising in the inner region of the accretion torus,
but in a region further out.

\subsection{Stellar Wind and Disk Wind Components of the Br$\gamma$ Emission}

The emission in the M17 sources is much less intense than for NGC3576
\#48, and this is consistent with the fact that the photospheres of
both sources have been observed at shorter wavelengths as well
\citep{hhc97}. The M17 sources are somewhat more evolved than BN,
MWC--349, and NGC3576 \#48. The central sources have had more time to
clear away the over--lying birth material. In the case of M17--268,
the observed Br$\gamma$ emission consists of two readily separable
components. An extended narrow component which appears roughly uniform
in extent (at least on scales of $\sim$ $\pm$ 5$''$) and a spatially
compact (i.e. centered on the continuum source) and very broad
component. The shape of the compact/broad Br$\gamma$ emission in
M17--268 appears as a double--peaked profile. The FWHM of the line is
approximately 480 \kms \ and the separation of the two peaks is $\sim$
176 \kms. The background subtracted emission of M17--275 is also
broad, but not double--peaked. This source exhibits broad wings with a
central component which is $\sim$ 130 \kms \ FWHM. The extent of the
wings is difficult to judge because of the rather small wavelength
coverage provided by a single Phoenix grating setting. Ideally, a
larger continuum region should be observed on either side of the
line. Comparison of different telluric standards suggests the large
scale fluctuations in the continuum should be less than about 3$\%$,
so the very broad wings evident in Figure~\ref{brg275} may indeed be
real.

The basic features seen in the compact Br$\gamma$ spectra of M17--268
and M17--275 are consistent with the CO emission profiles. The broad,
double--peaked emission in M17--268 (Figure~\ref{brg268}) corresponds
with the broad CO profile, suggesting a more edge--on
inclination. Similarly, the narrower prominent emission in M17--275
corresponds to the lower vsini determined from its CO profile. The
broad wings of M17--275 could be produced in a fast wind originating
at the star. The full extent is difficult to determine, but seems at
least to be $\pm$ 300 \kms and may actually be greater than the full
coverage of the spectrum, $\pm$ 750 \kms. The spectra of M17--268 and
NGC3576 \#48 may also show weak, but broad wings in
emission. \cite{hs86} could not easily fit the Br emission in MWC~349
to such a wind model, but they were attempting to fit the strong,
double--peaked profile (like the prominent shape in
Figure~\ref{brgs48}). Our long slit data with a narrow slit allows for
a better attempt at separating components of emission which are likely
coming from different spatial scales/mechanisms.

\begin{figure}
\plotone{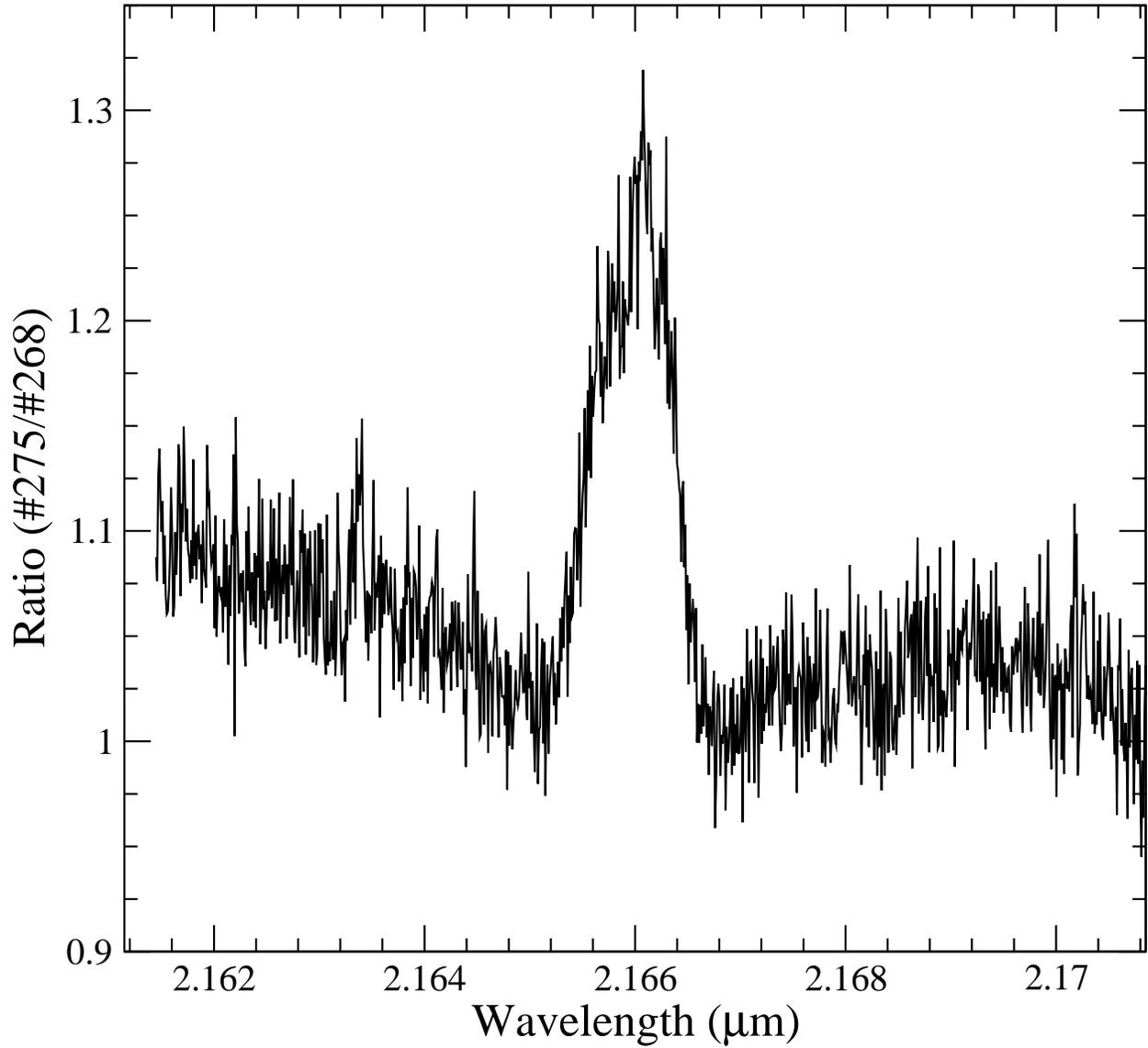} \figcaption{Ratio of background subtracted M17
sources (275/268). The increasing ratio in the wings of the spectra is
consistent with a larger velocity extent in \#275. This would be
expected if the wings are due to a fast wind since \#275 has a more
face--on geometry.\label{ratio}}
\end{figure}

If the broad wings in the M17 sources are due to a stellar wind, then
it is expected that the velocity extent seen in \#275 should be
greater than that seen in \#268 since the CO and prominent Br emission
suggest the latter has a more edge--on geometry. Figure~\ref{ratio}
suggests this may be the case. This figure shows a ratio of the \#275
to \#268 Br$\gamma$ spectra. The ratio tends to increase at the ends
of the spectrum as would be expected if the emission in the wings of
both sources is due to a wind.

\citet{hol93} and \citet{yw93} developed models for the
photoevaporation of accretion disks around massive stars in order to
explain the inferred long life times of ultra compact H~II (UCH~II)
regions based upon their numbers relative to the total numbers of OB
stars expected in the Galaxy \citep{wchwll89}. The accretion disk
could continually resupply fresh material as the outer layers of the
disk are ablated off due to the action of the ionizing continuum of
the central star and the ensuing heating of the disk gas; see the
recent review by \citet{hyj00}. \citet{kess97} modeled an 8.4
M$_\odot$ star and found that the predicted dust temperatures from
such a model do not match the temperatures derived from the observed
``unresolved'' type of UCH~II. Perhaps the observed properties of
UCH~II regions derive from more than one component (e.g a disk/torus
plus envelope). Many of the properties of NGC3576 \#48 and MWC~349 are
similar to UCH~II regions \citep[suggested that NGC3576 \#48 is an
UCH~II region]{barb03} and both exhibit emission line spectra
consistent with an ionized disk/torus. The M17 sources 268 and 275
also show spatially compact Br emission which is consistent with a
disk origin (especially 268).

In the model spectra produced by \citet{kess97}, the high velocity
wind included in their model is not evident. The lower density in the
wind and excitation conditions for the lines they chose to model
conspire to ``hide'' the wind compared to the more prominent emission
seen from the disk. Our observed Br spectra of NGC3576 \#48 show a
clear redshifted wing of higher velocity emission (and perhaps a blue
shifted wing as well), and the M17 sources both appear to exhibit very high
velocity wings. In all cases, the lower velocity emission is prominent
while the ``wind'' component is seen at lower intensity due to the
fact that it is highly dispersed. The spectrum of MWC~349 \citep{hs86}
was obtained through a much wider aperture than the slit used here
(3.75$''$) and this could further dilute the presence of a wind. The
nature of these wings should be investigated further with longer
wavelength coverage and detailed wind plus disk models.

\subsection{Timescales and the Accretion Process}

The ionizing radiation from massive stars, coupled with a stellar wind
is expected to have a profound effect on a circumstellar disk
\citep{hyj00}. \citet{ry97} calculated the disk dispersal times for
B--type stars (appropriate to the types of central stars discussed
here) from detailed hydrodynamical models. For a wide range of mass
loss properties, the dispersal time for early B--type stars with
ionizing continuum luminosities up to $\sim$ 10$^{48}$ s$^{-1}$, is
between 10$^5$ and a few 10$^6$ yr. 

We have carried out a survey of massive star clusters embedded
in giant H~II (GHII) regions in the Galaxy
\citep{bdc99,bcd00,bdc01,fig02,fig04}. Coupled with the work of
\citet{hhc97} in M17, several observations can be made about the
timescales for disk dispersal as a function of mass on the zero age
main sequence (ZAMS) and/or the formation mechanism for massive
stars. This survey covers the most luminous GHII in the Galaxy as
tabulated by \citet{s78}. Because massive star clusters will naturally
disperse their birth material as they evolve, these luminous clusters
must be young. 

While the survey is not yet complete (specifically, we have yet to
obtain near infrared spectra for the brightest members of all the GHII
stellar clusters), to--date we find that in each case where mid to
early O--type stars (i.e. the most massive stars) are revealed in the
cluster, only second rank late O or early B stars exhibit signatures
of circumstellar envelopes or disks/tori (W31, W43, W42, and M17 from
Hanson et al. 1997). In no cluster in the survey (nor any other
reported in the literature to our knowledge) has a young {\it massive}
O star been found which has the clear disk signatures seen in the
lower mass OB stars such as those presented here (i.e. the classic CO
emission profile). This may naturally be due to a shorter dispersal
time scale for the more energetic stars; on the other hand, it is also
consistent with a difference in formation mechanism for the most
massive stars--they may form by collisional processes instead of pure
accretion.

In the former case, the models of \citet{ry97} suggest the stars in
our sample are less than about 1 Myr old. The models and the observed
lack of massive O stars with similar disk signatures suggest the
dispersal timescale must be quite short, perhaps significantly less
than 10$^5$ yr. Clearly, higher mass detailed models would be very
useful. The near infrared observations may be approaching their limit
for massive stars. If the timescale for dispersal is so short, then
the prospect of observing the inner accretion disk in CO at these
wavelengths for massive O stars is diminished since the overlying
envelope may completely obscure the inner regions.

\section{SUMMARY}

We have presented high dispersion near infrared spectra of a sample of
massive stars still involved in the birth process and embedded within
massive star clusters. The spectra reveal CO 2.3 \mic \ 2--0
rotational--vibrational bandhead profiles which are well fit by a
rotating Keplarian disk for three of the four stars in the sample. For
the fourth, NGC3576 \#48 which exhibits a narrow CO profile, other
circumstellar geometries are possible. In particular, we present
evidence that the CO emission in this case may arise in a region
between the expanding cavity surrounding the central source and its
more compact accretion torus. Similar mechanisms may be at work in the
more well known BN and MWC~349 objects.

The CO profiles in three of our four stars provide solid evidence that
massive stars do indeed form by accretion. Together with similar CO
profiles presented by \citet{chan95} for two early B--type stars, it
is clear that stars as massive as 10--30 M$_\odot$ form with accretion
disks just as for lower mass stars. The timescale for disk dispersal
by the ionizing and wind action of massive stars must be quite short
for the most massive stars in Galactic young clusters because none has
been observed with a similar signature as that for the second rank OB
stars. Alternately, the formation mechanism for the most massive stars
may be different.

CLB and AD thank FAPESP and PRONEX for support. PSC appreciates
continuing support from the National Science Foundation. We appreciate 
the critique of an anonymous referee which has helped to improve our paper.

\newpage 


\clearpage
                                             
 
\begin{deluxetable}{lllccr}
\tablecaption{Table of Observations}
\tablehead{
\colhead{Object} &
\colhead{Date\tablenotemark{a}} &
\colhead{Line} &
\colhead{$K$} &
\colhead{$H-K$} &
\colhead{Reference\tablenotemark{b}} 
}
\startdata
NGC3576 \#48   & Feb 14, 2002; Feb 17, 2003 & CO             & 8.35 &2.21 &\citet{fig02} \\
NGC3576 \#48   & April 18, 2003             & Br$\gamma$, Br$\alpha$    &      &     &\citet{fig02} \\ 
M17 \#268      & May 11, 2003               & CO, Br$\gamma$ & 9.6  &1.0  &\citet{hhc97} \\
M17 \#275      & May 11, 2003               & CO, Br$\gamma$ & 8.0  &1.2  &\citet{hhc97} \\
G333.1-0.4 \#4 & July 29, 2003              & CO             &10.92 &1.44 &\citet{fig04} \\
\enddata

\tablenotetext{a}{UT Date of the observations.}

\tablenotetext{b}{Reference presenting object photometry, low
resolution spectra, object coordinates, and finding chart.}

\end{deluxetable}

\begin{deluxetable}{lccccccccc}
\tablecaption{CO Line Profile Parameters and Stellar Properties}
\tablewidth{0pt}
\rotate
\tabletypesize{\scriptsize}
\tablehead{
\colhead{Object} &
\colhead{$p$\tablenotemark{a}} &
\colhead{$q$\tablenotemark{b}} &
\colhead{$\beta$\tablenotemark{c}} &
\colhead{Model CO Line Width\tablenotemark{d}} &
\colhead{$N_{\rm CO}$\tablenotemark{e}} &
\colhead{vsini\tablenotemark{f}} &
\colhead{V$_\circ$\tablenotemark{g}} &
\colhead{Mass\tablenotemark{h}} &
\colhead{Luminosity} \\
\colhead{} &
\colhead{} &
\colhead{} &
\colhead{} &
\colhead{(\kms)} &
\colhead{(10$^{21}$ cm$^{-2}$)} &
\colhead{(\kms)} & 
\colhead{(\kms)} & 
\colhead{(M$_\odot$)}&
\colhead{(L$_\odot$)}\\ 
}
\startdata
NGC3576 \#48   &0.5 &1.0 $\pm$ 0.03 &2.1 $\pm$ 0.04 & 1.0 & 13.4 $\pm$ 0.1 &  25.0 $\pm$ 0.2 & $-$21.5 $\pm$ 1.3 \tablenotemark{i} &17\tablenotemark{j}&$>$50000\tablenotemark{j}\\
M17 \#268      &0.5 &3.8 $\pm$ 0.1  &5.8 $\pm$ 0.1  & 1.0 & 22.3 $\pm$ 1.8 & 258.7 $\pm$ 1.9 &    0.0  $\pm$ 1.3  &10&5000\\
M17 \#275      &0.5 &4.1 $\pm$ 0.1  &6.9 $\pm$ 0.3  & 1.0 &  3.5 $\pm$ 0.2 & 109.7 $\pm$ 0.6 & $-$7.0  $\pm$ 1.3  &15&20000\\
G333.1-0.4 \#4 &0.5 &3.6 $\pm$ 0.3  &6.5 $\pm$ 1.0  & 1.0 &  1.3 $\pm$ 0.4 & 108.0 $\pm$ 1.9 & $-$26.0 $\pm$ 1.4  &5\tablenotemark{j} & $>$300\tablenotemark{j}\\
\enddata

\tablenotetext{a}{Temperature power--law exponent where T $=$ T$_\circ$
(r/ri)$^{-p}$. T$_\circ$ $=$ 5000 K for all models and ri is the 
inner radius of CO emission zone.}

\tablenotetext{b}{Surface Density, $\Sigma$, power--law exponent where
$\Sigma$ $=$ $N_{\rm CO}$ (r/ri)$^{-q}$ and ri is the inner radius of CO
emission zone.}

\tablenotetext{c}{Ratio of the outer to inner radius of the CO emission zone.}

\tablenotetext{d}{The intrinsic line width at a point in the disk. The
models adopt Lorentzian profiles as in \citet{naj96}, but Gaussian
profiles of larger width can also fit the data.}

\tablenotetext{e}{Surface density at the inner edge of the CO emission
zone.}

\tablenotetext{f}{The projected rotational velocity of the Keplarian disk
at the inner edge of the CO emission zone.}

\tablenotetext{g}{Velocity shift from rest wavelength. Not corrected to 
heliocentric velocity.}

\tablenotetext{h}{Stellar mass derived from luminosity and associated
spectral type (see reference listed in Table~1). For the M17 sources,
the mass comes from the placement in the H--R diagram. For the
remaining sources, the mass is estimated from the ZAMS tabulation of
\citet{hhc97}.}

\tablenotetext{i}{The radio recombination line velocity of NGC3576 is
given as $-$24 \kms \ by \citet{depree99} and is consistent with the
velocity of the bulk of the extended Br$\gamma$ emission \#48.  The
mean observed CO velocity corresponds to a Heliocentric velocity of
$-$8.8 \kms. See also Table note $a$ of Table 3.}

\tablenotetext{j}{Approximate lower limit to the source luminosity and mass
derived by assuming an approximate maximum excess emission; see
reference quoted in Table~1.}

\end{deluxetable}

\begin{deluxetable}{lccc}
\tablecaption{Observed Br$\gamma$ Emission Velocities}
\tablewidth{0pt}
\tabletypesize{\scriptsize}
\tablehead{
  
  \colhead{Object} &
  \colhead{V$_\circ$\tablenotemark{a} (\kms)}&
  \colhead{FWHM\tablenotemark{b} (\kms)} &
  \colhead{P2P\tablenotemark{c} (\kms)} 
}
\startdata
NGC3576 \#48   & $-$66 $\pm$ 1 & 124 $\pm$ 5  & 84 $\pm$ 1 \\
M17 \#268      & 5 $\pm$ 2     & 477 $\pm$ 7  & 166 $\pm$ 6 \\
M17 \#275      & $-$16 $\pm$ 2 & 121 $\pm$ 3  & \\
\enddata

\tablenotetext{a}{Mean velocity of compact emission, e.g., the
``torus'' or ``disk wind'' component.Compare to values given in
Table~2 for the mean CO emission velocity. This is the observed
velocity. For \#48 the Heliocentric velocity is within 1 \kms \ of the
observed velocity. For the M17 sources the Br$\gamma$ and CO data were
obtained on the same night, so they are directly comparable. The
quoted errors are dominated by the continuum choice; see text.}

\tablenotetext{b}{Full width at half maximum for the same component as
$a$. The quoted errors are dominated by the continuum choice; see text}

\tablenotetext{c}{Peak--to--peak velocity separation for the torus or
disk wind components in NGC3576 \#48 (Figure~\ref{brgs48} and M17--268
(Figure~\ref{brg268}).}

\end{deluxetable}

\clearpage

\setcounter{page}{18}


.

.  

\end{document}